%Paper: astro-ph/9209001
%From: ROBIN%FROBES51.BITNET@vm.cnuce.cnr.it
%Date: 4 SEP 92 12:55:33.40-GMT
%Date (revised): 18 SEP 92 13:46:08.45-GMT

%\documentstyle[a4d,12pt]¤artdble‡
\documentstyle[a4,11pt]¤article‡
\def\sun¤\hbox¤$\odot$‡‡
\def\acknowledgements¤\vspace¤12pt‡\noindent¤\em Acknowledgements.\/ ‡%
\ignorespaces‡

\begin¤document‡
\setcounter¤page‡¤0‡
\title¤\bf The radial structure of the galactic disc ‡
\thanks¤Based on observations made at
Canada-France-Hawaii Telescope (CFHT)‡
\author¤\small Annie C. Robin$^¤1,2‡$, Michel Cr\'ez\'e$^¤2‡$,
Vijay Mohan$^¤3‡$\\
¤\small $^¤1‡$Observatoire de Besan\c¤c‡on, BP1615, F-25010 Besan\c¤c‡on
Cedex, France‡ \\
¤\small  $^¤2‡$CNRS URA1280, Observatoire de Strasbourg,
11 rue de l'Universit\'e,‡\\
¤\small F-68000 Strasbourg, France‡ \\
¤\small $^¤3‡$U.P. State Observatory, Manora Peak, Nainital, 263129 India ‡‡

\maketitle
\begin ¤abstract‡
Three colour photometry on CCD frames in the Special Area SpA23
provides a deep probe of the galactic disc in a low absorption
window
towards the anticenter. Magnitudes to better than 10\% at V = 25
and
B-V colour down to V = 23 have been obtained. These new data,
used in
combination with lower magnitude photographic data in a wider
field,
give a strong evidence that the galactic density scale length is
rather
short (2.5 kpc) and drops abruptly beyond 6
kpc.

\end¤abstract‡

\section ¤ Introduction ‡ Currently accepted ideas concerning the
structure and evolution of stellar populations in our Galaxy are
mainly
based on detailed observations of stars in the close solar
neighbourhood. The raw observational data which can be obtained
for many
faint stars do not allow deriving intrinsic stellar parameters such
as
distance, mass, age, evolutionary stage, chemical composition,
interstellar extinction of individual stars. However, some
information
relevant to the distribution of these quantities is reflected in the
n-dimensionnal distribution of observables. Magnitudes and colours
are
also connected to ages and star formation processes through the
history
of star formation and evolution. Connecting observable
distributions to
the main processes they come from is basically a multivariate
problem
for which we have developed at the Observatoire de Besan\c¤c‡on
both a
synthetic approach of Galaxy modelling (hereafter referred to as
the
Besan\c¤c‡on model, Robin and Cr\'ez\'e, 1986, Bienaym\'e et al.,
1987)
and a multivariate sample survey plan including complete samples
of UBV
photometry and proper motions in a set of galactic fields.

Mohan et al. (1988) performed UBV star counts on Schmidt plates
in the direction of the Special Selected Area (SpA23) near the
galactic
anticenter.
The selected field was identified by Kapteyn as a low extinction
window.
Mohan's observations are complete in U-B and B-V down to V
magnitude
16.5. This intermediate magnitude sample was shown to provide
tight
constraints for the disc scale length and the local extinction.
Going to deeper magnitudes gives the possibility to
probe the stellar populations further away from the Sun, giving
access to
the disc edge, and to test intrinsically fainter populations
(K and M dwarfs, white dwarfs).  This paper is dedicated to
processing
deeper probes in this direction (SpA23, $l=179^\circ, b=2.5^\circ$).
CCD
photometry is obtained in three photometric bands close to the
Johnson
UBV system down to magnitudes as faint as 25 in V.

\section ¤ CCD photometry ‡
\subsection ¤ Photometric system ‡ CCD observations have been
obtained with
the 3.6 meter Canada-France-Hawaii Telescope. We have used the
RCA2 CCD
along with the wide field corrector at the prime focus, giving a
scale
of 13.9"/mm. The dimensions of the chip are 640x1024 pixels,
which
correspond to about 8 square arcminutes. To improve the statistics
we have observed 4 neighbouring fields giving a survey area 29
arcmin$¤^2‡$. The UBV filters
have been selected to match as closely as possible the Johnson
system,
even at the expense of loss of sensitivity. The filters used were:

\indent V :  CFHT Mould V

\indent B :  CFHT B2 (BG12-2mm, BG18-2mm, GG385-2mm)

\indent U :  UG1 + CuSO4.

The B filter has been chosen so that, when multiplied by the
spectral
response of the CCD, it produced a passband as close as possible to
the
Johnson B filter. This caused a certain loss of transmittence, but
resulted in a colour coefficient close to unity. The colour equations
to
convert the instrumental system to the Johnson one have been
computed by
Mohan et al. (1988). They are:
\begin¤equation‡ V = v  \label¤eq:color1‡
\end¤equation‡
\begin¤equation‡ B-V = 1.06(b-v) - 0.68 \label¤eq:color2‡
\end¤equation‡
\begin¤equation‡ U-B = 1.36(u-b) - 1.92\label¤eq:color3‡
\end¤equation‡
\begin¤table*‡
\caption[]¤
 Field coordinates and exposure times in seconds in each
field.‡
\begin¤flushleft‡
\begin¤tabular‡¤|l|llllcccl|‡
\hline
Field  &  RA (1950) & Dec (1950) & l & b & V & B & U & Dates\\
\hline
1       &5 53 06.5&+30 39 43&   179.70 & 2.88 & 7200&
        6300    &13076 & jan 1987, jan 1988\\
2       &5 52 45.4&+30 40 52&   179.64 & 2.83 & 6000&
        6300    &10800 & jan 1988\\
3       &5 52 54.7&+30 36 13&   179.73 & 2.82 & 6000&
        7200    &10800 & jan 1988\\
4       &5 52 46.4&+30 38 24&   179.68 & 2.81 & 6000&   7200
        &10560 & jan 1988\\
\hline
\end¤tabular‡
\end¤flushleft‡
\end¤table*‡

\subsection ¤ Data reduction ‡
In each field a large number of short exposures have been taken
alternatively with standard field exposures, allowing to obtain an
absolute calibration and to measure the air mass corrections. Long
exposures have been recoordinated and summed. Long
exposure data are given in Table 1. We totalise  about 1H40mn in V,
2h in B and 3 to
3h30 in U in each field.

The image processing was carried out using the DAOPHOT software
package
(Stetson, 1987) in the ESO-MIDAS environment. First, stars were
detected
above a given threshold and stored in a catalogue of stars with
positions,
estimated magnitudes, roundness and sharpness according to the
mean
FWHM. Then aperture photometry was performed for each detected
star. A
point spread function (PSF) was computed by choosing isolated
stars in
the field. The number of PSF stars was of the order of 10 to 12 for
short
exposures and 15 to 25 for long exposures. Most of the time we
found
enough isolated stars such that we did not need to iterate the
process
(by subtracting neighbouring stars using the first guessed PSF and
recomputing it with
new isolated stars). Finally magnitudes were computed by fitting
the PSF
to each star through an iterative
process. The resulting magnitudes were then shifted to the
aperture
magnitude by a constant computed with the isolated stars.

The CCD magnitudes on short exposure frames were calibrated
using the
photometric sequence of the region S241 (Moffat et al., 1979) and
NGC2301 (Hoag et al., 1961). The S241 sequence was observed
several times during each night in alternance with fields 1 and 2 in
each filter and at different airmasses. The list of standards has
been given
in Mohan et al. (1988), table 9. Extinction slopes have been
computed to get
magnitudes outside the atmosphere. Fields 3 and 4 were calibrated
using
fields 1 and 2 as standards. Colour equations were then used to
transform
the magnitudes to the Johnson system.  Finally long exposures were
calibrated differentially with short exposures in each field. The
photometric accuracy has been estimated by comparison of the
magnitudes measured on various exposures. Table 2 gives the
estimated
accuracy as a function of magnitude down to the completeness limit in
each band. It should be noted that the
accuracy degrades at bright magnitudes because the PSF is
distorted and that at faint magnitudes the
degradation depends on the seeing. In the U band it seems that at faint
magnitudes stars are lost before the accuracy get less than 0.1.

In this reduction process
dedicated
to crowded fields, the galaxies are lost, while some are visible in
two
of our fields.

\begin¤table‡
\caption[]
¤Estimated photometric accuracy as a function of
magnitude in each band‡
\begin¤flushleft‡
\begin¤tabular‡¤|c|ccc|‡
\hline
 Magnitude  &  V  &  B  &  U\\
\hline
18-19   &0.02&0.02&0.02\\
19-20   &0.01&0.02&0.02\\
20-21   &0.02&0.01&0.01\\
21-22   &0.02&0.01&0.01\\
22-23   &0.04&0.03&0.01\\
23-24   &0.05&0.04&0.04\\
24-25   &0.08&0.09&   \\
\hline
\end¤tabular‡
\end¤flushleft‡
\end¤table‡

\subsection ¤ Detection limit and completeness limit ‡

Stars are detected to very faint magnitudes (28 in V) but they are
measured
with a very poor accuracy. They are not useful since their
brightness
is measured with an accuracy of the order of one
magnitude or more. For our purpose we need a catalogue with a very
well defined
completeness limit. Stars
are lost far before the detection limit because the signal to noise
ratio degrades rapidly. The completeness failure currently appears
in the
magnitude histogram of any passband 2 or 3 magnitudes below the
detection limit. The accuracy is better than 0.1 magnitude at 25 in
V.
To remain with a catalogue with a good accuracy we choose the
limiting
magnitude at the maximum of the histogram corresponding to an
estimated
error of about 0.1 magnitude. The completeness limits in each field
are given
in table 3.

\begin¤table‡
\caption[]
¤Field characteristics in each photometric band :  area in square
arcminutes, completeness limit and estimated seeing in
arcseconds‡
\begin¤flushleft‡
\begin¤tabular‡¤|cccccccc|‡
\hline
 Field & Area & \multicolumn¤2‡¤c‡¤V‡ & \multicolumn¤2‡¤c‡¤B‡ &
\multicolumn
¤2‡¤c|‡¤U‡ \\
 & & lim. & " & lim. & " & lim & " \\
\hline
1       &       6.7&    25      &1.9    &       22.5&1.7&22&1.7\\
2       &       6.5&    25      &1.8    &       24&1.0&22&1.3\\
3       &       7.7&    25      &1.3    &       24&1.4&21&1.3\\
4       &       6.6&    25      &1.6    &       23&1.6&23&1.5\\
\hline
\end¤tabular‡
\end¤flushleft‡
\end¤table‡

The catalogues of position, magnitude and colours in the Johnson
system
of the stellar objects brighter than 25 in V
are available at the Centre de Donnees astronomiques de Strasbourg (CDS)
or on request.

\section ¤ Model fitting ‡While a global analysis of all program
fields
is necessary to tune our model to reality, a field by field analysis
should be performed first to evaluate primary values of important
parameters.
Also, eye comparisons help in estimating the quality of model
predictions. In
the following we first determine the interstellar extinction in
this field from the U-B,B-V distribution, then we
deduce the density law characteristics of the stellar
populations towards the anticenter.

\subsection  ¤ (U-B, B-V) diagrams ‡ The number of stars in each
field
with a measured U-B colour index (about 40 to 50) allows to
estimate the
mean reddening for stars with V less than about 21 using colour-
colour
diagrams. Since the main sequence is nearly parallel to the
reddening
line in the U-B/B-V diagram, the measure of the reddening by
fitting a
shifted main sequence is rather inaccurate. We prefer to properly
model
the extinction using our model of population synthesis which will
be
used to interpret the stellar statistics. The model has been
extensively described in Robin, Cr\'ez\'e (1986) and in Bienaym\'e
et
al. (1987).

In the model simulated star counts, we try various density
distributions
of the absorbing material. An Einasto law (Einasto, 1977; see also
sect. 3.2 below) has been used to
model the absorbing layer as a disc quite similar to the youngest
stars'.
In addition to this smooth distribution absorbing clouds can be
introduced as
well along the line of sight.
The amount
of local differential absorption may be either tuned to the
data or estimated using maps  of interstellar matter. A value of 0.7
magnitude per kiloparsec is suitable for fields at intermediate and
high galactic
latitude.

Mohan et al. (1988) determined the absorption law in SpA23 field
from their Schmidt data. They found a
value of 0.7 magnitude per kiloparsec from stars of magnitude 12
to 16.
Since the bulk of CCD stars, ranging between 17 and 25 are on the
average more distant, the absorption law should be reinvestigated.
\begin¤table*‡
\caption[]
¤Estimation of the interstellar reddening in each magnitude bin
from the U-B/B-V diagrams to V=21. Mean distance and dispersion
of
subsample stars are given in columns 5 and 6. Column 7 is the
model predicted A$_¤V‡$ (best fit) at the mean distance.‡
\begin¤flushleft‡
\begin¤tabular‡¤|c|cccccc|‡
\hline
 Data & Magnitude & E(B-V) & A$_¤V‡$ & Mean d & $\sigma$(d) &
A$_¤V, model‡$\\
& Intervalle  &        &         &(kpc) &      &\\
\hline
Schmidt
&12-13          &0.2   & 0.6 &   1.2&  0.6 & 0.8\\
&14-15          &0.3   & 0.9 &  2.1&  1.3 & 1.1\\
&15-16          &0.33  & 1.0 &  2.6&  1.7 & 1.2\\
&16-16.5        &0.40  & 1.2 &  2.8&  2.0 & 1.2\\
\hline
CCD
&15-17          & 0.34 & 1.02 & 2.5 & 1.7 & 1.2\\
&17-19          & 0.35 & 1.06 & 3.3&  2.2 & 1.2\\
&19-21          & 0.45 & 1.35 & 4.1&  2.6 & 1.2\\
\hline
\end¤tabular‡
\end¤flushleft‡
\end¤table*‡

Using the same method we estimate the mean interstellar extinction in
the four fields simultaneously. To reduce the number of parameters, the
extinction law is assumed not to vary within the 5' by 5' joint fields.
Figure~1 shows a set of model predicted diagrams based on three values
of diffuse absorption from 0.6 to 1.0 magnitude per kpc together with
the unreddened main sequence. Matching simulated diagrams to observed
ones in different V magnitude intervals (visually and by computing a rms
estimation of the agreement) gives a best fit value of $0.8 \pm 0.1$
magnitude per kiloparsec for the local differential visual extinction
and A$_¤V‡$ = 1.2 $\pm$ 0.15 at 4 kpc. Figure~1d shows the observed
two-colour diagram of stars in the 4 CCD fields alongwith the unreddened
main sequence. Finally we give in Table 4 the mean distance and colour
excess of model stars in each magnitude bin, assuming that the ratio
between the visual absorption and the reddening is 3.0 as suitable for
dwarfs stars. These average points are plotted in fig.~2 over a set of
model absorption curves.

\subsection ¤Global extinction in the field‡

In the 4 CCD frames a number of galaxies are easily
distinguishable.
Among the brightest we identify 4 ellipticals which can help to
determine the overall extinction in the Galaxy since they have
quite
well defined intrinsic colors (de Vaucouleurs, 1976) B-V = 0.9
$\pm$ 0.1
and U-B = 0.45 $\pm$ 0.15.

The V magnitude and B-V colour have been measured for the 4
ellipticals
as well as
U-B for the brightest. We deduce the B-V excess according to
supposed intrinsic colours (table~5). The values are slightly
higher than the reddening determined from the UBV diagram but are
within
the measuring errors. Only one
galaxy shows a significantly higher extinction.

\begin¤table‡
\caption[]
¤Colours of ellipticals measured in the CCD fields. E(B-V) is the
reddening in front of the galaxies assuming intrinsic B-V of 0.9 .‡
\begin¤flushleft‡
\begin¤tabular‡¤|cc|ccccccc|‡
\hline
& field & Id & V & B-V & U-B & E(B-V) & $\sigma$ & A$_¤V‡$\\
\hline
& 2 & 1 & 18.40 & 1.42 & 0.66 & 0.52 & 0.10 & 1.56\\
& 2 & 2 & 21.46 & 1.46 & - & 0.56 & 0.15 & 1.68\\
& 4 & 3 & 21.78 & 1.62 & - & 0.72 & 0.15 & 2.16\\
& 4 & 4 & 22.12 & 1.49 & - & 0.59 & 0.15 & 1.77\\
\hline
\end¤tabular‡
\end¤flushleft‡
\end¤table‡

Previous estimations of the amount of absorption in anticenter
fields
close to ours led to similar results. SpA23 is at the limit between
two
zones investigated by Neckel and Klare (1980). They estimate the
total
visual extinction between 0.5 and 1.2 in the first zone, and
between 1.2 and
1.9 in the second one.
It is well in agreement with the presently
determined value of 1.2 $\pm $ 0.15.

McCuskey (1967) gives A$_¤V‡$ = 1.8 mag. at 2 kpc
and 2.3
at 4 kpc in a field at $l=186^\circ, b=+1^\circ$, significantly above
our
determination. However
the McCuskey's field is situated at 7 degrees in longitude from
SpA23 and is
closer to
the galactic plane. Deutschman et al. (1976) from the  Celescope
catalogue for
quite a
large region in Auriga-Gemini found  values between 1.2 and 1.8
mag. of visual
extinction
at 3 kpc. Moreover West (1967) when studying the galactic cluster
M37, located
at $l=178^\circ$ and $b= 3^\circ$, found an extinction of 0.8 mag.
at 1450 pc.
The present value of
extinction at 4 kpc of 1.2 $\pm$ 0.15 and differential absorption of
0.8 $\pm$ 0.1 magnitude
per kiloparsec determined
independently by UBV photographic photometry on Schmidt plates
and by
UBV CCD photometry is thus fully compatible with the extinction
measured
in neighbouring fields. This is in accordance with the fact that this
Kapteyn's Special Area was chosen as a zone of low extinction.

This value is also compatible with HI data in this direction which
give
a value of E(B-V)~=~0.4.

\subsection ¤ V, B-V data and deep V counts ‡ We now try to
constrain
the galactic structure parameters which are likely to produce a
significant signature in these star counts. Apart from the
extinction
which has been studied above, the density law of the stellar disc
must be
derived mainly from counts in V and B-V, because too few stars
have a
measured U-B index (not fainter than 21 in V). We describe the
density
law towards the anticenter by two parameters, the disc scale
length, hl,
and the maximum distance at which we see the stars, $r_¤max‡$ :

\[ \rho (r) = \left\¤\begin¤array‡¤ll‡
   K exp(0.5-\sqrt¤0.5^¤2‡+(a/hl)^¤2‡‡) & \mbox¤if $r<r_¤max‡$‡\\
   0. & \mbox¤otherwise‡
\end¤array‡
\right. \]

where $a = \sqrt¤R^¤2‡ + (z/c)^¤2‡‡$, R and z are the galactic
coordinates and c is the axis ratio function of the age of the
population, ranging from 0.02 to 0.073 in the disc. This formula
gives a
density equivalent to an exponential radially, while vertically it is
close to a sech$^2$  (see Bienaym\'e et al., 1987, for details).

In a previous study Mohan et al. (1988) determined the disc scale
length
from Schmidt plate photometry completed by a preliminary
reduction of
CCD data of our field I to magnitude 21. The Mohan data were shown
compatible only with a scale length of the order of $2.2 \pm 0.3$
kpc
which is the value used in our standard model. The deepest and
more
extensive CCD samples show a change of slope which could not be
detected at
magnitudes brighter than 21. No satisfactory fit can now be
obtained by
simply fitting the disc scale length and the extinction law. A
cutoff of
the disc has to be introduced.

In Figure 3 (a to f) the observed apparent magnitude distribution is compared
with
a series of model predictions with scale lengths ranging from 2.2
to 4.5 kpc. In
each figure the model is plotted without cutoff (dashed line) and
with
the cutoff value resulting from the maximum likelihood estimation
(solid line).

A grid of models with various scale lengths has been studied. For
each of
these models we determine the best value of
the cutoff using a maximum likelihood technique. Since the
magnitude
range 12 to 16 is only slightly sensitive to the disc cutoff (as we
see on
fig.~3) we use only the CCD data ranging between 16 to 25 in a
first
step to determine the cutoff. The likelihood for each model is
computed
as described in Bienaym\'e et al. (1987, appendix C) :

Let $q_¤i‡$ be the number of stars predicted by the model in bin $i$
and
$f_¤i‡$ be the  observed number. In case the deviations of  $f_¤i‡$'s
with
respect to  $q_¤i‡$ just reflect random fluctuations in counts,
each $f_¤i‡$ would be a Poisson variate with mean  $q_¤i‡$. Then
the
probability that $f_¤i‡$ be observed is :

\begin¤equation‡
dP¤_i‡ =   \frac¤q_¤i‡^¤f_¤i‡‡‡¤f_¤i‡!‡  exp ( - q_¤i‡)  \label¤eq:lik1‡
\end¤equation‡

Then the likelihood of a set of $q_¤i‡$'s given the relevant $f_¤i‡$
is :
\begin¤equation‡
L = ln \sum dP_¤i‡ = \sum_¤i‡ (- q_¤i‡ + f_¤i‡ \ln q_¤i‡ - \ln
f_¤i‡!)\label¤eq:lik2‡
\end¤equation‡

In search of the models that maximise L it is convenient to use the
reduced form :

\begin¤equation‡
L - L_¤0‡ = \sum_¤i‡ f_¤i‡ (1-\frac¤q_¤i‡‡¤f_¤i‡‡ +
ln\frac¤q_¤i‡‡¤f_¤i‡‡) \label¤eq:lik3‡
\end¤equation‡

where $L_¤0‡$ is constant and $L-L_¤0‡$ = 0 for a model which
would
exactly predict all $f_¤i‡$'s.

The likelihood computation is made in bins of V and B-V for stars
having
a B-V measure, that is for B less than 22.5 in all CCD fields. These
data
are binned with a step of 2 in V and 0.3 in B-V.
Below V=20 B-V data fail to be complete in the red wing (B is
complete to 22.5). Thus for some stars only V is available. To
constrain the model to V=25 we add in the likelihood computation
five bins of V magnitudes between 20 and 25 with a step of 1.
We end up with a
set of 23 bins in which the likelihood is computed. Table~6 gives
the
likelihood of the CCD data considering a grid of models with
different
scale lengths and disc
cutoffs. The disc cutoff is very well determined for long
scale lengths although in the case of short scale the maximum of
the
likelihood is wider (of course at very short scale lengths the model
predictions are cutoff insensitive). However the disc scale length
is not well
constrained by the CCD data alone.

\begin¤table*‡
\caption[]
¤likelihood of CCD data under various assumed model disc scale
length hl and cutoffs $r_¤max‡$.‡
\begin¤flushleft‡
\begin¤tabular‡¤|c|ccccccc|‡
\hline
 $r_¤max‡/ hl$ &  1.5 & 1.8 & 2.0 & 2.2 & 2.5 & 3.0 & 3.5\\
\hline
   1.00 &   -1474. &   -1187. &   -1324. &   -1316. &   -1171. &    -
912. &    -981.\\
   1.50 &   -1046. &    -747. &    -746. &    -895. &    -635. &    -509.
&    -488.\\
   2.00 &    -811. &    -602. &    -549. &    -533. &    -622. &    -469.
&    -363.\\
   2.50 &    -588. &    -407. &    -335. &    -278. &    -289. &    -191.
&    -161.\\
   3.00 &    -450. &    -271. &    -203. &    -156. &    -119. &     -83.
&     -56.\\
   3.50 &    -359. &    -186. &    -142. &     -82. &     -48. &     -54. &
-21.\\
   4.00 &    -284. &    -141. &    -104. &     -56. &     -21. &     -44. &
-52.\\
   4.50 &    -244. &    -106. &     -74. &     -37. &      -8. &     -49. &
-62.\\
   5.00 &    -213. &     -80. &     -51. &     -42. &     -33. &     -80. &
-152.\\
   5.50 &    -207. &     -73. &     -45. &     -34. &     -33. &    -100. &
-191.\\
   6.00 &    -190. &     -67. &     -34. &     -31. &     -26. &    -133. &
-275.\\
   6.50 &    -205. &     -59. &     -29. &     -27. &     -34. &    -183. &
-368.\\
   7.00 &    -200. &     -56. &     -25. &     -28. &     -43. &    -208. &
-425.\\
   7.50 &    -183. &     -48. &     -19. &     -34. &     -57. &    -264. &
-527.\\
   8.00 &    -180. &     -46. &     -19. &     -38. &     -66. &    -295. &
-580.\\
   8.50 &    -164. &     -44. &     -17. &     -40. &     -73. &    -323. &
-635.\\
   9.00 &    -161. &     -43. &     -17. &     -44. &     -81. &    -346. &
-678.\\
   9.50 &    -157. &     -40. &     -16. &     -48. &     -88. &    -374. &
-731.\\
  10.00 &    -153. &     -39. &     -16. &     -53. &    -108. &    -412. &
-822.\\
  10.50 &    -150. &     -39. &     -16. &     -55. &    -114. &    -433. &
-868.\\
\hline
\end¤tabular‡
\end¤flushleft‡
\end¤table*‡

The bulk of stars in Schmidt data lies at relatively short distances
providing a good basis to constrain the scale length within the
first
3~kpc. Binning the Schmidt data by step of 1 magnitude in V and by
0.3
in B-V, we apply the same likelihood technique to the whole set of
data
(Schmidt + CCD), giving a total of 58 bins. Table 7 gives the
resulting
likelihoods.

\begin¤table*‡
\caption[]
¤likelihood of Schmidt + CCD data under various assumed model
disc scale
lengths hl and cutoffs $r_¤max‡$.‡
\begin¤flushleft‡
\begin¤tabular‡¤|c|ccccccc|‡
\hline
 $r_¤max‡$ / hl &  1.5 & 1.8 & 2.0 & 2.2 & 2.5 & 3.0 & 3.5 \\
\hline
   1.00 &   -8678. &   -8168. &   -7237. &   -7493. &   -6958. &   -
5848. &   -5931.\\
   1.50 &   -6844. &   -5577. &   -4992. &   -5220. &   -4446. &   -
3672. &   -3616.\\
   2.00 &   -4968. &   -3808. &   -3679. &   -3412. &   -3022. &   -
2387. &   -2162.\\
   2.50 &   -3907. &   -3015. &   -2541. &   -2322. &   -2004. &   -
1516. &   -1598.\\
   3.00 &   -3387. &   -2368. &   -1941. &   -1730. &   -1231. &    -
987. &   -1127.\\
   3.50 &   -2847. &   -1870. &   -1427. &   -1110. &    -790. &    -
690. &    -840.\\
   4.00 &   -2605. &   -1624. &   -1182. &    -910. &    -607. &    -
611. &    -851.\\
   4.50 &   -2444. &   -1453. &   -1018. &    -748. &    -515. &    -
656. &    -975.\\
   5.00 &   -2264. &   -1264. &    -831. &    -636. &    -461. &    -
764. &   -1324.\\
   5.50 &   -2235. &   -1212. &    -786. &    -592. &    -444. &    -
816. &   -1459.\\
   6.00 &   -2210. &   -1139. &    -705. &    -529. &    -438. &    -
952. &   -1797.\\
   6.50 &   -2192. &   -1086. &    -683. &    -513. &    -492. &   -
1131. &   -2156.\\
   7.00 &   -2164. &   -1076. &    -671. &    -514. &    -521. &   -
1208. &   -2321.\\
   7.50 &   -2138. &   -1061. &    -663. &    -524. &    -557. &   -
1328. &   -2567.\\
   8.00 &   -2129. &   -1056. &    -660. &    -531. &    -571. &   -
1384. &   -2658.\\
   8.50 &   -2111. &   -1052. &    -658. &    -536. &    -583. &   -
1435. &   -2753.\\
   9.00 &   -2107. &   -1051. &    -659. &    -540. &    -596. &   -
1476. &   -2828.\\
   9.50 &   -2101. &   -1047. &    -657. &    -547. &    -608. &   -
1514. &   -2907.\\
  10.00 &   -2096. &   -1044. &    -657. &    -555. &    -633. &   -
1562. &   -3033.\\
  10.50 &   -2093. &   -1040. &    -658. &    -556. &    -640. &   -
1589. &   -3095.\\
\hline
\end¤tabular‡
\end¤flushleft‡
\end¤table*‡

The maximum likelihood is obtained with a scale length of 2.5 kpc,
and a
cutoff at 5.5 to 6 kpc from us. There however remains a
discrepancy
with the data in 8 bins in the best model. These discordant bins
indicate that there may be some discrepancy in the population
distribution in the HR diagram used in the present model. Since the
agreement is obtained in the large majority of the bins we expect
that
these spurious bins would not change the conclusion about the
space
density of stars obtained in the overall analysis.

\section ¤ Discussion ‡
\subsection ¤Validity of the fitting method ‡
The maximum likelihood is strictly valid only in so far as the space
of
possible models is sufficiently explored - which may not be the
case if
even a single important parameter has been overlooked. In
our modelling the basic ingredients
relevant to the interpretation of a galactic anticenter field are
(i) the
distribution of stars in the HR diagram (here mainly disc main
sequence
and few disc giants at $V<14$) - including their age distribution,
(ii) the global shape of the galactic disc (i.e. mainly the disc
scale length and cutoff),
(iii) the distribution of the absorbing layer.

The age distribution does not play an important role in the galactic
plane
since scale heights do not matter much in this low latitude field.
We assumed the
distribution of stars in the HR diagram to be roughly known from
solar
vicinity data, at least enough to be used as a standard candle to
probe
the external part of the galactic disc. We do not see in the data any
feature that might suggest a significant effect related to spiral
arms.

Under these assumptions the CCD data do provide a clear answer to
the
question of the disc shape : a satisfactory agreement cannot be
obtained
unless we impose a disc cutoff at something like 6 to 8 kpc away
from
the solar neighbourhood. It is remarkable that the fit to the CCD
data
is correct over the whole magnitude and colour range.

Once this best cutoff has been obtained for each model for the CCD
data,
the analysis is redone on the whole data set (CCD + Schmidt data).
The best model is still the same but the likelihood degrades
dramatically in some bins (which would be equivalent to a high
level of
rejection in a chi-square test). However the degradation is only due
to
a few bins - namely 8 bins out of 58, all in the Schmidt plate data.
But there is no way to change the shape of the Galaxy model so as
to get
a better fit consistently in all bins.
The same solution based only on three free parameters (scale
length, single cutoff and extinction) does fit all the CCD bins,
which
correspond to the better known part of the main sequence (F to K
dwarfs), and also the
large majority of the Schmidt plate photometry.

\subsection ¤Disc scale length‡

The resulting scale length of 2.5 kpc determined by our data is
smaller than what
is currently accepted for the galactic disc. Values obtained in the
visible are of the order of 3.5 to 4.5 kpc but there are estimations
ranging
between
1.8 to 6 kpc. Kent et al. (1991) give a review of recent
determinations. 2.2 micron determinations give quite short scale
lengths (Maihara et al. (1978): 1.8 kpc, Jones et al. (1981): 2.0 kpc,
Eaton et al. (1984): 3.0 kpc, Kent et al. (1991): 3.0 kpc). IRAS OH/IR
star data give larger values : Habing (1988): 4.2 kpc, Rowan-
Robinson
and Chester(1987): 6.0 kpc. Visible data show a large spread as
well : De Vaucouleurs and Pence (1978): 3.5 kpc, Lewis and Freeman
(1989): 4.4 kpc, van der Kruit (1986): 5.5 kpc, Oblak and Mayor
(1987) :
2.5 kpc. Most such measurements are model dependent and require
assumptions.

The Pioneer 10 mission has been used to map the background
starlight of
the Galaxy in two wavelength bands: ``blue'' at 395-485 nm and
``red'' at
590-690 nm.
A scale length of 5.5 kpc has been deduced from the Pioneer data
by van der Kruit (1986) through a number of assumptions.
The Bahcall and Soneira model (1984) was used to deconvolve data
since
only integrated surface brightness over the line of sight was
available.
The comparison between model and data led to major problems
(model
too bright at high latitudes, too big slope in the profile and too
faint
at intermediate latitudes) and a rescaling of the local luminosity
density
of the model was necessary to fit the data. It is difficult to
estimate
the effect of such a rescaling on the resulting scale length.
Moreover,
Pioneer data do not constrain directly the scale length but the
ratio
$\frac¤hl‡¤hz‡$ such that the results depend on the assumed scale
height.
They emphasized a value of 325 pc (that is the value of old disk
dwarfs)
while one expects that old disk giants would dominate the light in
Pioneer data (Lewis and Freeman, 1989).
Lewis and Freeman (1989) determined the radial scale length from
the
kinematics of disk K giants. They found a different scale for
$\sigma_¤R‡$ and for $\sigma_¤\phi‡$ (4.4 and 3.4 kpc
respectively).
These values should depend on assumptions on the axisymmetry of
the
Galaxy.

The measurement of the asymmetric drift gives an alternative to
determine the radial scale length. It depends on the sum of the
radial density gradient and the velocity dispersion gradient
$\frac¤\partial ln \rho‡¤\partial R‡ + \frac¤\partial ln
\sigma_¤U‡^¤2‡‡¤\partial R‡$.

Mayor (1974) found -0.65 kpc$^¤-1‡$ for the total value using a
sample of
1010 A and F stars with uvby photometry. The value of the velocity
dispersion gradient may be determined using the stability criterion
of
the galactic disc introduced by Toomre (1964). In a Schmidt model
(1965)
Mayor deduced a value of -0.2 for $\frac¤\partial ln \sigma_¤U‡^¤2‡‡
¤\partial R‡$ in contrast to the value of 0 that one would have in
the
ellipsoidal hypothesis. This result has been confirmed by Oblak and
Mayor (1987) from F, G and K type stars in the Gliese catalogue
who
found values between -0.14 and -0.24. It is also in good agreement
with
the estimation of Vandervoort (1975) and Erickson (1975) based on
the
3rd and 4th order moments of the local stellar velocity
distribution
(between $-0.19$ and $-0.23$ kpc $^¤-1‡$).

The radial density gradient $\frac¤\partial ln \rho‡¤\partial R‡$
implied by these values is $-0.45$ kpc
$^¤-1‡$
corresponding to a scale length of 2.2 kpc, in perfect agreement
with our
measurement.

\subsection ¤The edge of the galactic disc‡

The disc cutoff has been found at a quite short distance (6 kpc),
giving a galactic radius of 14 to 15 kpc. Habing
(1988) found a cutoff at even shorter distances (between 1 and 2.5
kpc)
from the study of OH/IR stars from the analysis of the IRAS Point
Source
Catalogue. However the model he used was based on a disc scale
length of
4.5 kpc, quite larger than ours. Moreover Habing
mentionned that these parameters were not strongly constrained
by his
data and that a number of slightly different models would also be
compatible with these IRAS data. It would be interesting to verify
that
our present measures of the disc scale length and cutoff are
compatible
with Habing's analysis.

While our measurement is compatible with a rather sharp cutoff at
the
edge of the disc it might be slightly smoother according to the
population
studied. Remote open clusters have been found at quite large
distances
in the anticenter (Be 21 at 14.5 kpc from the center and To 2 at 16
kpc,
Janes, 1991) which is not incompatible with our result.

\subsection ¤Star distribution in the HR diagram ‡ Among the 58
bins
where there are stars in the (V,B-V) grid, 8 bins show a significant
difference between the number of stars predicted and the observed
one.
The interpretation that comes out is that the frequency of stars in
the parts of the
HR diagram which correspond to these bins is not realistic. They
correspond to 4 zones:

(i) $12<V<14$ and $B-V<0.3$ : These stars are A and F stars of
absolute
magnitude 0. to 1., and too many are produced by the model. This
may call into question the adopted dwarf to giant ratio in this part
of the HR diagram computed from the Michigan catalogue.

(ii) $12<V<13$ and $0.6<B-V<0.9$ : They are G subgiants lacking in
the model
by a factor of 2. Their absolute magnitude has been adopted to be 3.
There are some suggestions that this magnitude could be slightly
too bright
(Grenon, private communication). Since they have been computed
from local counts the inferred local density depends strongly on
the assumed absolute magnitude (see below). A further study of the
subgiant absolute magnitude and local density would clarify this
point.

(iii) $15<V<16.5$ and $0.3<B-V<0.6$ : These stars are F stars of
absolute
magnitude 1 to 3.5 which are lacking in the model. This point may
be related to the dwarf to subgiant ratio (see above).

(iv) $14<V<16.5$ and $1.2<B-V<1.5$ : Too many K giants are
produced by the
model in these bins. This problem has already been addressed
(Robin,
1989). We showed that the absolute magnitudes of K giants are
very
uncertain because for a given spectral type a large dispersion in
absolute magnitude is in fact observed (the so called `funnel
effect'). The mean magnitude obtained
is very much dependent on the selected sample.

In our approach the number of giants (and subgiants) predicted is
very sensitive to the
assumed absolute magnitude since we recompute their local
density from
the distribution of these stars in the solar neighbourood. An
error of 0.5 magnitude on $M_¤V‡$ implies an error of a factor 2 on
the
sampling volume, hence on the computed local
density. Our present estimate of absolute
magnitude is good to about 0.5 magnitude only. We need to modelise
in
more detail this part of the HR diagram (accounting in detail for
the
mixing of stars of various ages and masses) to be able to fit
perfectly
those star counts. The Hipparcos mission would be able to partly
solve
this problem - at least for the estimation of the absolute
magnitudes.
However, the problem of mixing stars of different types in the
same part of the HR
diagram will remain, until we have better stellar evolution models.

However, we hope that while studying in detail the (V,B-V)
distribution of the stars, we will be able to constrain the stellar
distribution in the HR diagram with a better accuracy than is
presently
achievable.

\section ¤ Conclusion ‡
We have combined star count data towards the anticenter using
Schmidt
Plates and CCD frames to constrain a model of galatic structure.
Thus we probe a magnitude range from 10 to 25 in V, with
two colour indices for stars with $V \leq 21$ and one colour index
to
magnitude 23. We are able to determine the extinction law from
the (V,U-B,B-V)
data cube and to deduce the disc scale length
and cutoff. In the anticenter direction the disc shows quite a short
scale length of 2.5 kpc and the stellar density drops abruptly
beyond 6
kpc.

This drop cannot be explained by the existence of a dense cloud
causing interstellar extinction
because the U-B,B-V stellar data, as well as the colours of 4
elliptical
galaxies in the field, exclude a large amount of extinction.

Stars counted on CCD frames should have absolute luminosities
between 4
and 13 (M$_¤V‡$) while those dominating Schmidt plate counts
range
between 0 and 5. A complete ad hoc revision of the luminosity function from 0
to 13 would be necessary in order to get a reasonnable fit over the whole V
distribution.
So there is no realistic change of the luminosity
function that might produce a fit over the whole magnitude
range.\\

\acknowledgements¤This research was partially supported by the
Indo-French
Centre for the Promotion of Advanced Research / Centre Franco-
Indien
Pour la Promotion de la Recherche Avanc\'ee.‡

\begin¤thebibliography‡¤‡
\bibitem¤ba‡
Bahcall, J.N., Soneira, R.M. 1984, ApJS, 55, 67
\bibitem
Bienaym\'e, O. Robin, A.C., Cr\'ez\'e, M. 1987, A\&A, 180, 94
\bibitem¤de:da‡
Deutschman, W.A., Davis, R.J., Schild, R.E. 1976, ApJS, 30, 97
\bibitem¤ei‡
de Vaucouleurs, G. 1976, Le Monde des Galaxies, A. Hayli (ed),
Observatoire de Besan\c¤c‡on
\bibitem¤de:pe‡
de Vaucouleurs G., Pence, W.D. 1978 , AJ, 83, 1163
\bibitem¤ea‡
Eaton, N, Adams, D.J., Giles, A.B. 1984, MNRAS, 208, 241
\bibitem¤ei‡
Einasto, J. 1977, IAU Symp. 84, The large Scale Characteristics of
the
Galaxy, ed. W.B. Burton, p. 451
\bibitem¤er‡
Erickson, R.R. 1975, ApJ, 195, 343
\bibitem¤ha‡
Habing, H.J. 1988, A\&A, 200, 40
\bibitem¤ho:jo‡
Hoag, A.A., Johnson, H.L., Iriarte, B., Mitchell, R.I., Hallman,
K.L., Sharpless, S. 1961, Publ. U.S. Nav. Obs., 17, part 7
\bibitem¤ja‡
Janes, K.A. 1991, Precision Photometry, A.G.D. Philip, A.R. Upgren
and
K.A. Janes (eds), L. Davis Press, p. 233
\bibitem¤jo‡
Jones, T.J., Ashley, M., Hyland, A.R., Ruelas-Mayorga, A. 1981,
MNRAS,
97, 413
\bibitem¤kd‡
Kent, S.M., Dame, T.M., Fazio, G. 1991, ApJ, 378, 131
\bibitem¤lf‡
Lewis J.R., Freeman, K.C. 1989, AJ, 97, 139
\bibitem¤mai‡
Maihara, T., Oda, N., Sugiyama, T., Okuda, H. 1978, PASJ, 30, 1
\bibitem¤may‡
Mayor, M. 1974, A\&A, 32, 321
\bibitem¤mc‡
McCuskey, S.W. 1967, AJ, 72, 1199
\bibitem¤mo:fi‡
Moffat, A.F.J., Fitzgerald, M.P., Jackson, P.D. 1979, A\&AS, 38, 197
\bibitem¤mo:bi‡
Mohan, V., Bijaoui, A., Cr\'ez\'e, M., Robin, A.C. 1988, A\&AS, 73, 85
\bibitem¤ne:kl‡
Neckel, Th., Klare, G. 1980, A\&AS, 42, 251
\bibitem¤om‡
Oblak, E., Mayor, M. 1987, Evolution of Galaxies, X IAU European
Meeting, J. Palous (ed), Publ. Astron. Inst. Czech. Acad. Sci. 69, 263
\bibitem¤ro:cr‡
Robin, A.C., Cr\'ez\'e, M. 1986, A\&A, 157, 71
\bibitem¤ro‡
Robin, A.C., 1989 A\&A, 225, 69
\bibitem¤rr‡
Rowan-Robinson, M. Chester, T. 1987, ApJ, 313, 413
\bibitem¤st‡
Schmidt, M.: 1965, Stars and Stellar Systems, Vol. V, Ed. A. Blaauw
and
M. Schmidt, Univ. of Chicago Press, p. 513
\bibitem¤st‡
Stetson, P.B. 1987, PASP, 99, 191
\bibitem¤to‡
Toomre, A. 1964, ApJ, 139, 1217
\bibitem¤va‡
Vandervoort, P.O. 1975, ApJ, 195, 333
\bibitem¤vdk‡
van der Kruit, P.C. 1986, A\&A, 157, 230
\bibitem¤we‡
West, F.R. 1967, ApJS, 14, 359
\end¤thebibliography‡
\newpage
\noindent ¤ \large \bf Figure captions‡\\\\
\noindent ¤\bf Figure 1:‡ Model simulated U-B vs B-V
diagram reddened by different values of local
differential absorption. a) 0.6 magnitude per kiloparsec. b) 0.8. c)
1.0. The reference sequence (solid line) is the unreddened main
sequence.
d) Observed diagram plotted over the
unreddened main sequence (solid line). A visual guide has been
printed
as a parallelogram where most of the stars are.\\\\

\noindent ¤\bf Figure 2:‡ Distribution of interstellar extinction along
the line of sight. Estimated values from (U-B, B-V) diagrams from CCD
(full circles), from Schmidt plates (triangles), and model with
differential local absorption  between 0.6 and 0.9 mag.~kpc$^¤-1‡$ (thin
lines).\\\\

\noindent ¤\bf Figure 3:‡ Distribution in V. Schmidt data (circles),
CCD data  (diamonds), models with various scale lengths : hl=2.0 (a),
2.2 (b), 2.5 (c), 3.0 (d), 3.5 (e), 4.5 (f). Solid line: model density
laws with best fit cutoff (9.5, 6.5, 4.5, 4.0, 3.5, 3.0 kpc resp.);
dashed line: model density laws assuming no cutoff.
\end